# Impact of Naturalistic Field Acoustic Environments on Forensic Text-independent Speaker Verification System

*Zhenyu Wang, John H.L. Hansen*
*Center for Robust Speech Systems (CRSS), University of Texas at Dallas, U.S.A*
`{zhenyu.wang, john.hansen}@utdallas.edu`

Audio analysis for forensic speaker verification offers unique challenges in system performance due in part to data collected in naturalistic field acoustic environments where location/scenario uncertainty is common in the forensic data collection process. Forensic speech data as potential evidence can be obtained in random naturalistic environments resulting in variable data quality. Speech samples may include variability due to vocal efforts such as yelling over 911 emergency calls, whereas others might be whisper or situational stressed voice in a field location or interview room (Hansen et al. 2015, Campbell et al. 2009) [1,2]. Such speech variability consists of intrinsic and extrinsic characteristics and makes forensic speaker verification a complicated and daunting task. Extrinsic properties include recording equipment such as microphone type and placement, ambient noise, room configuration including reverberation, and other environmental scenario-based issues (Hansen et al. 2015) [1]. Some factors, such as noise and non-target speech, will impact verification system performance by their mere presence. To investigate the impact of field acoustic environments, we performed a speaker verification study based on the CRSS-Forensic corpus (Wang et al. 2020) [3] with audio collected from 8 field locations including police interview. This investigation includes an analysis of the impact of seven unseen acoustic environments on speaker verification system performance using an x-Vector system.

A subset of the CRSS-Forensic corpus consisting of read speech, spontaneous speech and prompted speech from 75 speakers (27 male, 48 female), which was used for the experiments. This subset contains speech recorded in different acoustic environments such as a sound booth (33.9h) and public environments(99.4h). Speech data in the sound booth is recorded using a participant body-worn mobile data collection platform called LENA (sample rate: 16 kHz) as well as 6 microphones data recording platforms. In public environments, speech is collected by a LENA worn by the participant in seven recording environments. Additionally, referring to (Hansen et al. 1995) [4], a subjective score of stationarity(i.e., 1: for wide sense stationary, to 10:nonstationary) was assigned to each noise source field location based on first and second order moment analysis. The following seven indoor and outdoor locations are included with stationarity scores assigned [(1) to (10)]: *office* (1), *hallway* (3), *lobby* (3), *outside walking path* (5), *parking lot* (5), *cafeteria* (9), *public open game room* (9).

An x-Vector with PLDA (Probabilistic Linear Discriminant Analysis) based system (Synder et al. 2018) [5] was used for speaker verification experiments. Training data from the VoxCeleb corpus (Chung et al. 2018) [6] is augmented with noise, music, and babble speech from the MUSAN corpus (Synder et al. 2015) [7], with reverberation introduced using the RIR NOISES3 corpus. The augmented data consists of 7323 speakers with a total of 2.2M utterances. The model architecture consists of 5 time-delay network layers followed by a statistical pooling layer, two fully-connected layers with 512 units in each layer, and a probability output layer. We extract 30-dimensional MFCC features using a frame width of 25ms and a window shift of 10ms. For better speaker verification results over naturalistic field environments, we fine-tuned the parameters for the last time-delay layer before statistics pooling, and the first fully-connected layer after pooling in the pre-trained x-Vector model with our CRSS-Forensic data. Finally, the last fully-connected layer is replaced according to available speaker labels from CRSS-Forensic data.

For forensic analysis of the acoustic field environments, we adopted spectrogram for analysis. Fig. 1 visualizes speech (top plots) and ambient noise (bottom plots) time sequences of 5 sec duration for all locations (four representative samples shown here). The *office* is a relatively quiet environment, but sporadic overlapping speech is a main reason for lower system performance. *Hallway* and *lobby* locations are public indoor areas which contain sporadic non-target speakers and random ambient noise. The *walking path* and *parking lot* are two public outdoor areas, but *parking lot* is more quiet with fewer pedestrian conversations. The most noisy scenes are public university *cafeteria* and *game room*, where the target-speaker speech is often mixed together with other conversations. As shown in Fig 1 for *game room*, the ambient noise is strong enough to obscure the target speech. Accordingly, formants in the spectrum of *game room* speech are blurred, which means data is less discriminative for speaker identity or other forensic characteristics. The speech and noise energy in a 5 sec segment are quantified with mean value and standard deviation of the spectral amplitude response.

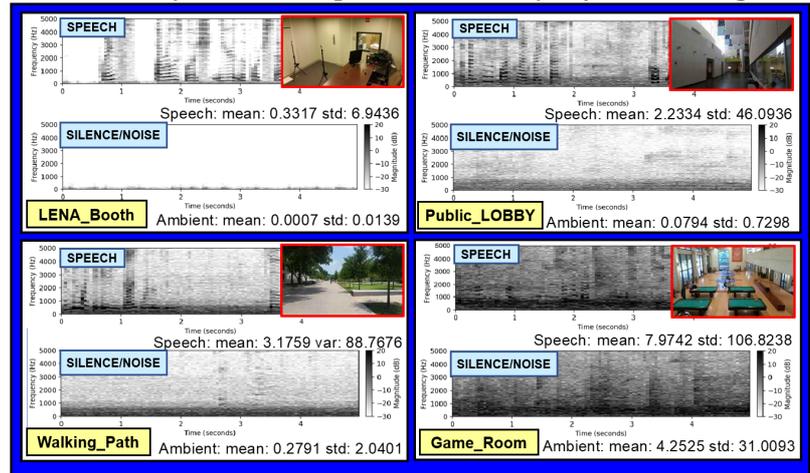

**Figure 1.** Spectrograms for different locations

To assess speaker verification system performance, we use the Detection Error Trade-off (DET) curve (Martin et al. 1997) [8] to evaluate systematic verification performance. Fig. 2 shows the Equal Error Rates (EER) for diverse field acoustic environments, where speaker verification system performance varies across different recording environments. Near microphone data recorded in the booth is taken as clean speech, which shows best verification results with an EER of 3.61%.

In this study, we have shown that forensic audio data collected from diverse naturalistic field acoustic environments often contain noise in varied stationarity levels, which impacts speaker verification system performance. The field locations with high non-stationarity acoustic conditions tend to obscure the target speaker identity information more. This also results in a low signal-to-noise ratio (SNR), which reduces the accuracy of speaker verification in naturalistic field locations.

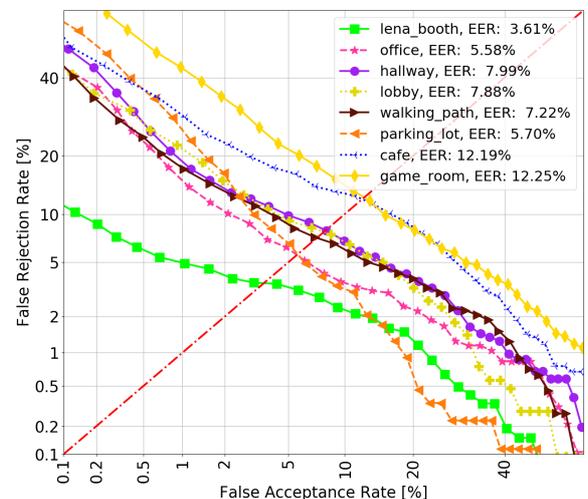

**Figure 2.** DET plot for each location